# Lassoing Line Outages in the Smart Power Grid

Hao Zhu and Georgios B. Giannakis

University of Minnesota, Dept. of ECE, 200 Union Street SE, Minneapolis, MN 55455, USA

*Abstract*—Fast and accurate unveiling of power line outages is of paramount importance not only for preventing faults that may lead to blackouts, but also for routine monitoring and control tasks of the smart grid, including state estimation and optimal power flow. Existing approaches are either challenged by the *combinatorial complexity* issues involved, and are thus limited to identifying single- and double-line outages; or, they invoke less pragmatic assumptions such as *conditionally independent* phasor angle measurements available across the grid. Using only a subset of voltage phasor angle data, the present paper develops a near real-time algorithm for identifying multiple line outages at the affordable complexity of solving a quadratic program via block coordinate descent iterations. The novel approach relies on reformulating the DC linear power flow model as a *sparse overcomplete expansion*, and leveraging contemporary advances in compressive sampling and variable selection using the least-absolute shrinkage and selection operator (Lasso). Analysis and simulated tests on the standard IEEE 118-bus system confirm the effectiveness of lassoing line changes in the smart power grid.

## I. INTRODUCTION

It is well appreciated that major blackouts have occurred partly due to lack of comprehensive situational awareness of the power grid [15]. This speaks for the importance of timely monitoring the status of generators, transformers, and transmission lines. Identifying outages and generally changes of lines is particularly critical for a number of tasks, including state estimation, optimal power flow, real-time contingency analysis, and thus security assessment of power systems.

To appreciate the opportunities and challenges facing the line-change identification problem, it is prudent to think of the grid as a graph comprising topologically interconnected power systems. Phasor measurement units (PMUs) provide voltage and power data per local (a.k.a. *internal*) system in real time. Likewise, real-time data are telemetered internally per system to offer topology-bearing information on the connectivity status of local circuit breakers and switches. On the other hand, power flow conservation across interconnected systems allows for identifying changes even in *external* system lines – a critical task for comprehensive monitoring. This would have been a non-issue if inter-system data were available. Unfortunately, the system data exchange (SDX) module of the North-American Electric Reliability Corporation (NERC) can provide grid-wide interarea *basecase* topology information on an hourly basis [16], but the desiderata is near real-time monitoring of transmission lines. In a nutshell, the need arises for each internal system to identify in a computationally efficient manner line outages (and generally monitor line changes) in its external counterpart relying only on basecase topology information and local PMU data.

Existing approaches typically formulate external line-outage identification as a combinatorially complex (integer programming) problem, which can be computationally tractable only for single- or at most double-line outages [5], [11], [12]. An alternative approach adopts a Gauss-Markov graphical model of the power network and can cope with multiple outages at affordable complexity, but assumes conditionally independent phasor angle measurements and requires inter-system PMU data to be available in real time across the grid [8].

The present paper contributes a computationally efficient algorithm for near real-time identification of *multiple* external line outages (and generally changes) using only hourly basecase topology information and *local* voltage phasor angle data available by PMUs. It relies on the standard DC linear power flow model, but also applies readily to its AC linearized counterpart per iteration. The novel approach views the topology-bearing basecase information as the weighted Laplacian matrix of the grid-induced graph (Section II). This leads to an overcomplete model of the changes-induced innovation, which in turn enables casting line-change identification as a *sparse* vector estimation problem (Section III). Solving the latter draws from recent advances in compressive sampling and variable selection in linear regression problems [1], [4], [13], and leverages the least-absolute shrinkage and selection operator (Lasso) to identify line changes at affordable complexity using provably convergent block coordinate descent (BCD) iterations (Section IV). Simulated tests corroborate the merits of the novel algorithm (Section V), and the paper is wrapped up with a concluding summary (Section VI).

*Notation:* Upper (lower) boldface letters will be used for matrices (column vectors); $(\cdot)^T$ denotes transposition; $(\cdot)^\dagger$ the matrix pseudo inverse; $\mathbf{1}$ the all-one vector; $\mathbf{0}$ the all-zero vector; $\|\cdot\|_p$ the vector $p-$norm for $p \geq 1$; $|\cdot|$ the magnitude or cardinality of a set; $\mathcal{S}_1 \backslash \mathcal{S}_2$ the relative complement of the set $\mathcal{S}_2$ in the set $\mathcal{S}_1$.

## II. MODELING AND PROBLEM STATEMENT

Consider a power transmission network consisting of $N$ buses denoted by the set of nodes $\mathcal{N} := \{1, \ldots, N\}$, and $L$ transmission lines represented by the set of edges $\mathcal{E} := \{(m,n)\} \subseteq \mathcal{N} \times \mathcal{N}$. Collect all the voltage phasor angles $\{\theta_n\}$, one per bus $n \in \mathcal{N}$, in the vector $\boldsymbol{\theta} \in \mathbb{R}^N$; and correspondingly the injected power variables $\{P_n\}$ in $\mathbf{p}$. The network

This work is supported by the NSF grants CCF-0830480, CCF-1016605, ECCS-0824007, and ECCS-1002180.

buses comprise the union $\mathcal{N} = \mathcal{N}_I \bigcup \mathcal{N}_E$ with $\mathcal{N}_I \bigcap \mathcal{N}_E = \varnothing$, where $\mathcal{N}_I$ denotes the subset of observable buses in the internal system, and $\mathcal{N}_E$ stands for the unobservable buses of the external system. Accordingly, $\boldsymbol{\theta}$ comprises two sub-vectors $\boldsymbol{\theta}_I$ and $\boldsymbol{\theta}_E$, collecting phasor angles of voltages measured at $\mathcal{N}_I$ and $\mathcal{N}_E$ buses, respectively; and similarly for the $\mathbf{p}_I$ and $\mathbf{p}_E$ partitions of $\mathbf{p}$. Supposing that the network-wide injected-power vector $\mathbf{p}$ only changes gradually across time, the present work aims to unveil line changes, including (possibly multiple) line outages anywhere in the network, using data $\boldsymbol{\theta}_I$ acquired in real time from PMUs. Before explicitly formulating the line-change identification problem, it is important to understand how the network topology dictates the relationship between $\mathbf{p}$ and $\boldsymbol{\theta}$, as detailed in the following.

### A. Linear DC Power Flow Model

Power flow models are useful for determining how injected power $\mathbf{p}$ flows along all transmission lines. To cope with the nonlinear AC power-voltage relationships, the DC power flow model offers a simple linear analysis tool, assuming constant voltage magnitudes per bus and negligible transmission losses [18, Sec. 4.1]. Although confined only to linear approximate analysis of the actual nonlinear system, the DC power flow model turns out to facilitate a variety of power system monitoring tasks under normal operating conditions, including security-constrained contingency analysis, and system state estimation; see e.g., [18, Ch. 11-12].

In the linear DC model, the power flowing from bus $m$ to bus $n$ through their connecting transmission line is given by

$$P_{mn} = \frac{1}{x_{mn}}(\theta_m - \theta_n), \quad \forall\, (m,n) \in \mathcal{E} \tag{1}$$

where $x_{mn} = x_{nm}$ denotes the reactance along line $(m,n)$. Notice that $P_{mn}$ can be either positive or negative, depending on whether actual power flows from bus $m$ to bus $n$, or in the reverse direction. Flow conservation implies that the power injected to bus $n$ balances all the line flows originating from it; that is,

$$P_n = \sum_{m \in \mathcal{N}_n} P_{nm} = \sum_{m \in \mathcal{N}_n} \frac{1}{x_{nm}}(\theta_n - \theta_m) \tag{2}$$

where $\mathcal{N}_n$ denotes the set of neighboring buses linked to bus $n$. Writing (2) in vector-matrix form yields

$$\mathbf{p} = \mathbf{B}\boldsymbol{\theta} \tag{3}$$

where the $N \times N$ matrix $\mathbf{B}$ has its $(m,n)$-th entry given by

$$B_{mn} = \begin{cases} -1/(x_{mn}) & \text{if } (m,n) \in \mathcal{E} \\ \sum_{\nu \in \mathcal{N}_n} 1/(x_{n\nu}), & \text{if } m = n \\ 0, & \text{otherwise.} \end{cases} \tag{4}$$

Matrix $\mathbf{B}$, relating the voltage-phasor angle $\boldsymbol{\theta}$ to the injected power $\mathbf{p}$ as in (3), is uniquely determined by the line reactance parameters $\{x_{mn}\}$ of the network $(\mathcal{N}, \mathcal{E})$, and topology-bearing information provided by the SDX. At this point, it is worth noting that each $x_{mn}$ reactance is present in only four entries of $\mathbf{B}$, namely, $(m,m), (m,n), (n,m)$, and $(n,n)$.

This observation about line's $(m,n)$ presence in $\mathbf{B}$ intuitively describes the selective effect topology changes due to line outages can have on the angle vector $\boldsymbol{\theta}$ in (3). To formulate this topological effect concretely, $\mathbf{B}$ can be viewed as the weighted Laplacian matrix of the graph $(\mathcal{N}, \mathcal{E})$. To this end, consider the $N \times L$ bus-line incidence matrix $\mathbf{M}$, see e.g., [17, pg. 56], formed by columns $\{\mathbf{m}_\ell\}_{\ell=1}^L$ of length $N$. With subscript $\ell$ corresponding to the line $(m,n)$, the column $\mathbf{m}_\ell$ has all its entries equal to $0$ except the $m$-th and $n$-th, which take the value $1$ and $-1$, respectively. The weighted graph Laplacian leads to the following representation of the network topology matrix as

$$\mathbf{B} = \mathbf{M}\mathbf{D}_x\mathbf{M}^T = \sum_{\ell=1}^L \frac{1}{x_\ell}\mathbf{m}_\ell\mathbf{m}_\ell^T \tag{5}$$

where the diagonal matrix $\mathbf{D}_x$ has its $\ell$-th diagonal entry $1/x_\ell$ equal to the inverse reactance $1/x_{mn}$, if $\ell$ corresponds to the line $(m,n)$. In addition, matrix $\mathbf{B}$ is symmetric, and has rank $N-1$ if the power network is connected, since its null space is only spanned by $\mathbf{1}$; see e.g., [17, pg. 469]. Rank deficiency of $\mathbf{B}$ gives rise to multiple solutions for $\boldsymbol{\theta}$ in (3). To fix this ambiguity, one generation bus is typically chosen as reference with its phasor angle set to zero, case in which phasor angles of all other buses denote their differences relative to the reference bus; see e.g., [18, pg. 76].

### B. Unveiling Network Faults of Line Outages

Given the linear DC power flow model (3), and the aforementioned reference bus convention, the pre-event phasor angles $\boldsymbol{\theta}$ can be uniquely characterized by the injected power vector $\mathbf{p}$ and the topology-dependent matrix $\mathbf{B}$. Suppose that due to changes in the grid, e.g., cascading failures in an early stage, several outages occur on lines, collected in the subset $\tilde{\mathcal{E}} \subseteq \mathcal{E}$. Line outages on the transmission network yield the post-event graph $(\mathcal{N}, \mathcal{E}')$, with[1] $\mathcal{E}' := \mathcal{E} \backslash \tilde{\mathcal{E}}$.

Without loss of generality (wlog) it is assumed that fast system dynamics are well damped, and that the system settles down to a quasi-stable state following the line outages. The possibility of poor system damping can be accounted for by low-pass filtering which smooths out system oscillations, as detailed in [11]. Under these considerations, the linear DC model for the quasi-stable post-event network is given by

$$\mathbf{p}' = \mathbf{p} + \mathbf{v} = \mathbf{B}'\boldsymbol{\theta}' \tag{6}$$

where $\mathbf{B}'$ is the post-event weighted Laplacian of $(\mathcal{N}, \mathcal{E}')$, and noise vector $\mathbf{v}$ accounts for the small perturbations between pre- and post-event injected power vectors due to, e.g., variations in the bus loads. Typically, bus power injections can be modeled as independent Gaussian-distributed random variables [10], since each bus load comprises a large number of individual users connected through their local distribution network. Thus, $\mathbf{v}$ can be reasonably assumed zero-mean Gaussian distributed, with covariance matrix $\sigma_v^2 \mathbf{I}$.

---

[1]As a mnemonic, post-event quantities are denoted with prime, and the differences relative to their pre-event counterparts are denoted with tilde.

Voltage phasor angles are only available at the subset of buses $\mathcal{N}_I$. Thus, unveiling outages amounts to identifying the lines in $\tilde{\mathcal{E}}$, given pre- and post-event phasor angle $\boldsymbol{\theta}_I$ and $\boldsymbol{\theta}'_I$. It turns out that solving this problem incurs combinatorial complexity. Using (3) and (6), the ensuing section will pursue an overcomplete representation for line outages. But before this, a remark is due on comparing the linear DC power flow model here with the probabilistic dependence graph in [8].

**Remark 1** (*Comparison with [8]*). To capture the spatial correlation of phasor angle data across the grid, a Gaussian Markov random field (GMRF) approach is pursued in [8] to model the statistical dependence of phasor angles among buses. Conditioned on $\{\theta_m\}_{m \in \mathcal{N}_n}$ at neighboring buses, the GMRF model in [8] assumes that $\theta_n$ is Gaussian distributed and independent of all other entries in $\boldsymbol{\theta}$. Unfortunately, the GMRF model falls short in capturing the actual power system operation for the following reasons. First, it assumes (conditionally) Gaussian distributed phasor angles. Second, it is not clear why conditional independence holds in practice. From the flow model (1), the power conservation law, and with $\{\theta_m\}_{m \in \mathcal{N}_n}$ given, the phasor angle $\theta_n$ is correlated with those of their two-hop neighbors. Lastly, as mentioned in Section I, it is currently impossible to acquire $\boldsymbol{\theta}_E$ in real time. In contrast, similar to [5], [11] and [12], the present paper adopts the linear DC flow model, which has been used extensively for various power system monitoring tasks.

## III. OVERCOMPLETE REPRESENTATION OF LINE OUTAGES

In this section, the line outage identification task is formulated as a sparse representation problem. To this end, the difference $\tilde{\mathbf{B}} := \mathbf{B} - \mathbf{B}'$ denoting the weighted Laplacian for the outage lines in $\tilde{\mathcal{E}}$, is written as [cf. (5)]

$$\tilde{\mathbf{B}} = \sum_{\ell \in \tilde{\mathcal{E}}} \frac{1}{x_\ell} \mathbf{m}_\ell \mathbf{m}_\ell^T. \quad (7)$$

Substituting the pre-event power flow model (3) into the post-event one in (6) leads to

$$\mathbf{B}\boldsymbol{\theta} + \mathbf{v} = \mathbf{B}'\boldsymbol{\theta}' = \mathbf{B}\boldsymbol{\theta}' - \tilde{\mathbf{B}}\boldsymbol{\theta}'. \quad (8)$$

Consider now the phasor angle change vector $\tilde{\boldsymbol{\theta}} := \boldsymbol{\theta}' - \boldsymbol{\theta}$, and partition it into the available $\tilde{\boldsymbol{\theta}}_I$ and the unavailable $\tilde{\boldsymbol{\theta}}_E$, each corresponding to buses in $\mathcal{N}_I$ and $\mathcal{N}_E$; and likewise, partition columns of $\mathbf{B}$ correspondingly to $\mathbf{B}_I$ and $\mathbf{B}_E$. With these notations, substituting (7) into (8), yields

$$\mathbf{B}\tilde{\boldsymbol{\theta}} = \mathbf{B}_I\tilde{\boldsymbol{\theta}}_I + \mathbf{B}_E\tilde{\boldsymbol{\theta}}_E = \tilde{\mathbf{B}}\boldsymbol{\theta}' + \mathbf{v} = \sum_{\ell \in \tilde{\mathcal{E}}} s_\ell \mathbf{m}_\ell + \mathbf{v} \quad (9)$$

where $s_\ell := \mathbf{m}_\ell^T \boldsymbol{\theta}'/x_\ell$, $\forall \ell \in \tilde{\mathcal{E}}$. To identify $\mathcal{E}'$ containing a given number of line outages, one can exhaustively test over all possible line combinations and select the one offering the least-squares (LS) fit of (9). Such an approach incurs combinatorial complexity, and has confined existing methods based on extensive enumeration of all combinations, to identifying single-line outage [5], [11], or at most double-line outages after reducing the search space [12].

To bypass this combinatorial complexity, the fresh idea here is to consider an overcomplete representation of all line outages. With the available $\mathbf{y} := \mathbf{B}_I \tilde{\boldsymbol{\theta}}_I$ from buses in $\mathcal{N}_I$, the data model in (9) can be reduced to a sparse linear regression by introducing the $L \times 1$ vector $\mathbf{s}$, whose $\ell$-th entry equals $s_\ell$, if $\ell \in \tilde{\mathcal{E}}$, and 0 otherwise. Thus, (9) can be written as

$$\mathbf{y} = \sum_{\ell \in \tilde{\mathcal{E}}} s_\ell \mathbf{m}_\ell - \mathbf{B}_E \tilde{\boldsymbol{\theta}}_E + \mathbf{v} = \mathbf{M}\mathbf{s} - \mathbf{B}_E \tilde{\boldsymbol{\theta}}_E + \mathbf{v} \quad (10)$$

where the incidence matrix $\mathbf{M}$, now viewed as a regression matrix, captures all the pre-event transmission lines. Introducing $\mathbf{M}$ allows the possibility of multiple line outages. Different from (9), the line outage set $\tilde{\mathcal{E}}$ is no longer present in (10), which eliminates the need for enumeration over all possible line combinations.

In essence, the overcomplete representation in (10) allows one to cast the problem of recovering the subset $\tilde{\mathcal{E}}$ as one of estimating the unknown angle vector $\tilde{\boldsymbol{\theta}}_E$ and the *sparse* vector $\mathbf{s}$. The key *premise* is that the number of line outages is a small fraction of the total number of lines; i.e., $|\tilde{\mathcal{E}}| \ll L$. This holds even for multiple line outages such as those occurring during cascading failures, at least in the early stage when only a small number of lines start to fail. The same premise is adopted for contingency analysis, where usually single- and double-line outages are of primary concern; see e.g., [18, Ch. 11]. Under this sparsity constraint, the underlying outage line locations are few, meaning that the vector $\mathbf{s}$ has only a few nonzero entries. In turn, this suggests recovering line outages by regularizing the LS criterion with the $\ell_1$-norm of $\mathbf{s}$, which promotes sparse solutions. Specifically, the pertinent objective becomes

$$\{\hat{\mathbf{s}}^\lambda, \hat{\boldsymbol{\theta}}_E^\lambda\} := \arg\min_{\mathbf{s}, \tilde{\boldsymbol{\theta}}_E} \|\mathbf{y} - \mathbf{M}\mathbf{s} + \mathbf{B}_E \tilde{\boldsymbol{\theta}}_E\|_2^2 + \lambda \|\mathbf{s}\|_1 \quad (11)$$

where $\lambda$ is a regularization parameter, and its choice will be detailed later on.

Using the overcomplete representation in (10), the line outage recovery problem has been formulated as the convex quadratic program (QP) in (11). Hence, the global minimizer of (11) can be efficiently obtained using general-purpose convex solvers, such as interior-point algorithms [3, Ch. 11]. Instead of these off-the-shelf solvers, the block coordinate descent (BCD) method will be adapted in the next section for solving (11), because it exploits the specific problem structure and can find the line outage identification path for a sequence of regularization parameters. Compared to previous works constrained to at most double-line outages [5], [11], [12], the proposed formulation overcomes the inherent combinatorial complexity with a virtual "outage" at every possible transmission line. By leveraging the sparse signal representation framework, the proposed approach offers the potential to achieve high accuracy in recovering faults with more than two line outages, while bypassing the computationally prohibitive enumeration of combinations for potential line outages.

**Remark 2** (*Partial information on* $\mathbf{v}$). The internal system formed by buses in $\mathcal{N}_I$ can also acquire $\mathbf{p}_I$ in real time via PMU data and state estimates [5]. In this case, the internal system has also available part of the noise vector $\mathbf{v}_I$ at the internal buses $\mathcal{N}_I$, and can thus exploit this information when estimating the vectors $\mathbf{s}$ and $\tilde{\boldsymbol{\theta}}_I$ in (10). Compared to (11), use of this additional information identifies line outages by minimizing the $\ell_1$-norm regularized LS cost for only part of the system in (10) corresponding to the buses in $\mathcal{N}_E$. With $\mathbf{v}_I$ available, strict equality constraints can be enforced on the remaining system corresponding to buses in $\mathcal{N}_I$. As expected, this extra information enhances accuracy and improves identifiability without sacrificing computational efficiency, since the problem is still a QP. Given this similarity, the remainder of the paper does not consider this extra information on $\mathbf{v}_I$.

**Remark 3** (*General models and line faults*). The DC power flow model has broad applicability in practice, since it also coincides with the system obtained after linearizing the decoupled *nonlinear* AC power flow model, see e.g., [2, Sec. 5.7]. Thus, the proposed overcomplete representation for line outages offers the potential to generalize the nonlinear AC power flow model via iterative approximation. Additionally, the expression of $\tilde{\mathbf{B}}$ in terms of the line reactance parameters in (7) suggests that the present analysis is also applicable to more general *line changes*. For example, if the $\ell$-th line experiences a sudden change in its reactance, then this can be also captured by $\tilde{\mathbf{B}}$ as differences in the diagonal weights for the Laplacian matrices $\mathbf{B}$ and $\mathbf{B}'$. In this sense, the line outage is just a special type of line change, since equivalently the line reactance goes to infinity, and its corresponding weight on matrix $\mathbf{B}'$ becomes zero. Therefore, the idea of a sparse overcomplete representation $\mathbf{s}$ of line outages readily extends to the *nonlinear* power flow model, and can be used to unveil general line parameter *changes*.

## IV. Line-Outage Identification Path via BCD

Although the QP problem in (11) is efficiently solvable using general-purpose convex optimization routines, tuning up the regularization parameter $\lambda$ is an issue to be addressed. In this section, first the identification path of $\hat{\mathbf{s}}^\lambda$ in (11) entailing line-outage vector estimates with variable sparsity levels is obtained using the block coordinate descent (BCD) method [14]. Subsequently, options are offered for selecting $\lambda$. The so-termed regularization path of Lasso-based solutions as a function of $\lambda$ has been studied extensively in the context of generalized linear model fitting; see e.g., [6]. This approach is advocated especially for sparse unknown vectors of high dimension. Consider for now (11) with a fixed regularization parameter $\lambda$. The BCD method optimizes the regularized LS cost in (11) by cyclically minimizing over the coordinates, namely, the vector $\tilde{\boldsymbol{\theta}}_E$, and all the scalar entries of $\mathbf{s}$. It yields successive estimates of one coordinate, while the rest are fixed. With the iterate $\mathbf{s}(i-1)$ for $\mathbf{s}$ available at the $(i-1)$-st iteration, $\forall\, i \geq 0$, the iterate $\tilde{\boldsymbol{\theta}}_E(i)$ of $\tilde{\boldsymbol{\theta}}_E$ can be estimated by solving the LS problem as

$$\tilde{\boldsymbol{\theta}}_E(i) = \arg\min_{\tilde{\boldsymbol{\theta}}_E} \left\| \mathbf{y} - \mathbf{M}\mathbf{s}(i-1) + \mathbf{B}_E \tilde{\boldsymbol{\theta}}_E \right\|_2^2$$
$$= -(\mathbf{B}_E)^\dagger [\mathbf{y} - \mathbf{M}\mathbf{s}(i-1)]. \quad (12)$$

Notice that the pseudo inverse of $\mathbf{B}_E$ needs only to be computed once for all iterations. Once $\tilde{\boldsymbol{\theta}}_E(i)$ becomes available, it remains to update the scalar entries of $\mathbf{s}(i)$ for iteration $i$. Suppose that the $\ell$-th entry $s_\ell(i)$ is to be found. Precursor entries $\{s_1(i), \ldots, s_{\ell-1}(i)\}$ have been already obtained in the $i$-th iteration along with $\tilde{\boldsymbol{\theta}}_E(i)$ obtained in closed form as in (12), and postcursor entries $\{s_{\ell+1}(i-1), \ldots, s_L(i-1)\}$ are also available from the previous $(i-1)$-st iteration. Thus, the effect of these given entries can be removed from $\mathbf{y}$ by forming

$$\mathbf{e}_\ell(i) := \mathbf{y} + \mathbf{B}_E \tilde{\boldsymbol{\theta}}_E(i) - \sum_{j=1}^{\ell-1} \mathbf{m}_j s_j(i) - \sum_{j=\ell+1}^{L} \mathbf{m}_j s_j(i-1). \quad (13)$$

Using (13), the vector optimization problem in (11) reduces to the following scalar one with $s_\ell(i)$ as unknown: $s_\ell(i) = \arg\min_{s_\ell}[\|\mathbf{e}_\ell(i) - \mathbf{m}_\ell s_\ell\|_2^2 + \lambda|s_\ell|]$. This is known to be the *scalar* Lasso problem, which admits a closed-form solution expressed via a soft thresholding operator as (see e.g., [6])

$$s_\ell(i) = \text{sign}(\mathbf{m}_\ell^T \mathbf{e}_\ell(i))\left[\frac{|\mathbf{m}_\ell^T \mathbf{e}_\ell(i)|}{\|\mathbf{m}_\ell\|_2^2} - \frac{\lambda}{2\|\mathbf{m}_\ell\|_2^2}\right]_+, \ell = 1, \ldots, L \quad (14)$$

where $\text{sign}(\cdot)$ denotes the sign operator, and $[\chi]_+ := \chi$, if $\chi > 0$, and zero otherwise.

Cycling through the closed forms (12)-(14) explains why BCD here is faster than, and thus preferable over general-purpose convex optimization solvers of (11). The LS regression in (12) involves only matrix multiplication, while the soft thresholding operation in (14) is even faster since the vector $\mathbf{m}_\ell$ has only two non-zero entries. Further, the BCD iteration is provably convergent to the global optimum $\{\hat{\mathbf{s}}^\lambda, \hat{\boldsymbol{\theta}}_E^\lambda\}$ in (11), as asserted in the following proposition.

**Proposition 1:** *(Convergence of BCD) Given the parameter $\lambda$ and arbitrary initialization, the iterates $\{\tilde{\boldsymbol{\theta}}_E(i), \mathbf{s}(i)\}$ given by (12) and (14) converge monotonically to the global optimum $\{\hat{\mathbf{s}}^\lambda, \hat{\boldsymbol{\theta}}_E^\lambda\}$ of the line outage identification problem in (11).*

*Proof:* The argument relies on the basic convergence result in [14]. It is easy to check that the two summand terms in the cost of (11) satisfy the assumptions (B1)–(B3) and (C2) in [14]. Convergence of the iterates $\{\tilde{\boldsymbol{\theta}}_E(i), \mathbf{s}(i)\}$ to a stationary point thus follows by appealing to both [14, Thm. 5.1] and the regularity property of the cost function [14, Lemma 3.1]. Convexity of the problem (11) further ensures that every stationary point is indeed the global optimum. Monotonicity of the convergence follows simply because the cost per iteration is strictly non-increasing. ∎

The identification path is further obtained by applying the BCD method for a monotonically decreasing sequence of $\lambda$ values. Larger values of $\lambda$ in (14) force more entries of $\mathbf{s}(i)$ to be shrunk to zero. Hence, if a large enough parameter $\lambda$ in (11) is picked, the corresponding $\hat{\mathbf{s}}^\lambda$ will eventually become zero. Further, with a decreasing sequence of $\lambda$ values, the optimal solution for a large $\lambda$ can be used as a *warm start* for solving (11) with the second largest $\lambda$. This way, the line outage identification path via BCD exploits both the efficient scalar solution in (14) as well as warm starts, to reduce complexity and improve algorithmic stability. The BCD based line-outage solution path of (11) is tabulated as Algorithm 1.

---

**Algorithm 1**: Input $\mathbf{y}$, $\mathbf{M}$, $\tilde{\mathbf{B}}_E$, and a decreasing sequence of $\lambda$ values. Output $\{\hat{\mathbf{s}}^\lambda, \hat{\boldsymbol{\theta}}_E^\lambda\}$ in (11) for each $\lambda$.

    Initialize with $i = -1$ and $\mathbf{s}(-1) = \mathbf{0}$.
    **for** each $\lambda$ value from the decreasing sequence **do**
      **repeat**
        Set $i := i + 1$ and update the iterate $\tilde{\boldsymbol{\theta}}_E(i)$ as in (12).
        **for** $\ell = 1, \ldots, L$ **do**
          Compute the residual $\mathbf{e}_\ell(i)$ as in (13).
          Update the scalar $s_\ell(i)$ via (14).
        **end for**
      **until** BCD convergence is achieved.
      Save $\hat{\boldsymbol{\theta}}_E^\lambda = \tilde{\boldsymbol{\theta}}_E(i)$ and $\hat{\mathbf{s}}^\lambda = \mathbf{s}(i)$ for the current $\lambda$ value.
      Initialize with $i = -1$ and the warm start $\mathbf{s}(-1) = \mathbf{s}(i)$.
    **end for**

---

### A. Tuning the Regularization Parameter

Algorithm 1 yields the line-outage identification path for a sequence of $\lambda$ values. Selection of $\lambda$ is a critical issue since a larger $\lambda$ promotes a sparser $\hat{\mathbf{s}}^\lambda$, which translates to a smaller number of line outages. With additional prior information, existing statistical tests can be adapted to select $\lambda$ corresponding to the actual number of line outages.

***Number of line outages is fixed or upper bounded***. By direct inspection of the regularization paths one can determine the value of $\lambda$, so that the degree of sparsity in $\hat{\mathbf{s}}^\lambda$ equals the number of line outages. Thus, the nonzero entries in the solution indicate the corresponding outage lines in $\tilde{\mathcal{E}}$. Note that the number of line outages is also assumed known in most existing works [5], [11], [12].

If the maximum number of line outages is prescribed, the identification path can yield the set $\tilde{\mathcal{E}}$ with all possible cardinalities less than the maximum. In order to determine the actual cardinality, one can adapt a minimum description length (MDL) type test; see e.g., [7]. Specifically, the rating score for a given number of line outages, becomes the sum of the LS reconstruction error achievable with the $\tilde{\mathcal{E}}$ of the specified cardinality, plus a penalty term linearly growing with the number of line outages. The "best" number of line outages, and thus $\tilde{\mathcal{E}}$, is then selected as the one yielding the minimum rating score.

***Variance of injected power noise is known***. If the variance $\sigma_v^2$ of the entries in $\mathbf{v}$ is known, one can proceed as follows. For each cardinality of the set $\tilde{\mathcal{E}}$, find the subset of buses which are not related to those outage lines. The effect of $\mathbf{Ms}$ at the corresponding entries of $\mathbf{y}$ in (10) then becomes zero. The LS reconstruction error from these entries of $\mathbf{y}$ can be calculated to obtain the sample variance of $\mathbf{v}$, as an estimate of $\sigma_v^2$. This test selects the set $\tilde{\mathcal{E}}$ with the minimum score, in terms of the sample variance deviation relative to the given $\sigma_v^2$.

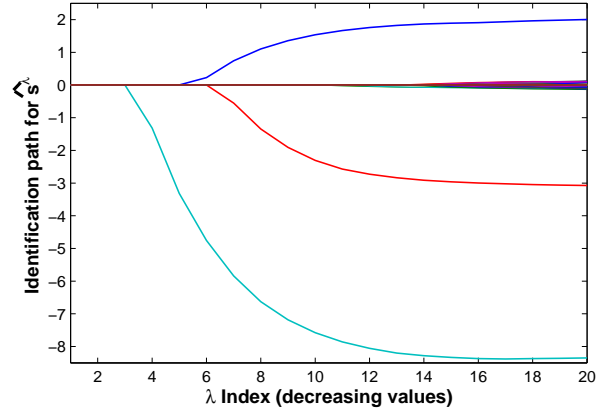

Fig. 1. Line-change identification path for $\hat{\mathbf{s}}^\lambda$ with a decreasing sequence of $\lambda$ values.

### V. NUMERICAL TESTS

The proposed sparsity-promoting fault diagnosis algorithm is tested in this section using the IEEE 118-bus benchmark system [9]. As in [5], the network is partitioned into internal and external systems $\mathcal{N}_I := \{1 - 45, 113, 114, 115, 117\}$ and $\mathcal{N}_E := \{46 - 112, 116, 118\}$, respectively. Also, the standard deviation for the noise $\mathbf{v}$ is set to be $5\%$ of one entry of the pre-event injected power vector $\mathbf{p}$.

The software toolbox MATPOWER [19] is used to generate the phasor angle measurements as well as the pertinent power flows. The set of outaged branches $\tilde{\mathcal{E}} := \{(42, 49), (38, 65), (69, 75)\}$ contains the lines indexed by $\ell = 66, 95, 115$, and is excluded from solving the post-event power flow. Notice that all these three outaged lines connect to buses in the unobservable external system $\mathcal{N}_E$, and especially the line $(69, 75)$ links two buses with unobservable measurements.

Algorithm 1 is applied to obtain the line outage identification path for $\hat{\mathbf{s}}^\lambda$ with an exponentially decreasing sequence of 20 $\lambda$ values, as depicted in Fig. 1. Clearly, a larger $\lambda$ yields fewer nonzero entries in $\hat{\mathbf{s}}^\lambda$, and thus leads to less line outages. In order to find the best number of line outages, the tests based both the MDL criterion and the sample variance deviation are implemented for $|\mathcal{E}'| = 1$ to $5$ line outages along the identification path. The MDL test scores follow from the one for general linear regression problems in [7], while the variance deviation scores are computed as the absolute differences between the sample and actual variances. The test scores for different numbers of line outages are scaled relative to the maximum ones in either test, and are plotted in Fig. 2. The test results demonstrate that the minimum scores are achieved at the actual number of line outages $|\mathcal{E}'| = 3$. Fig. 3 illustrates the absolution value of entries in $\hat{\mathbf{s}}^\lambda$, corresponding to a $\lambda$ value that yields three line outages. All entries in Fig. 3 are numerically equal to zero, except for three entries, coinciding exactly with the lines in $\mathcal{E}'$. All simulation results manifest the effectiveness of the proposed algorithm, in terms of unveiling multiple line outages and inferring topology changes at unobservable external systems.

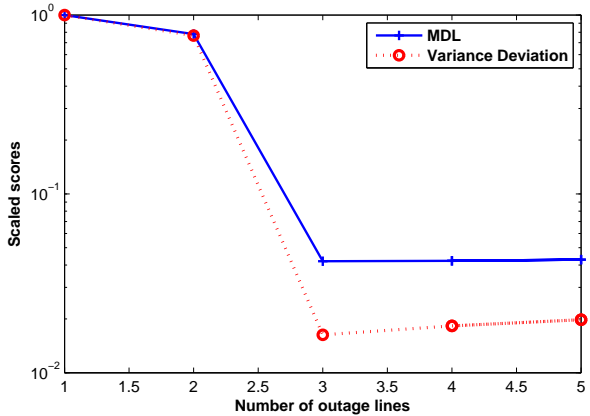

Fig. 2. Scaled scores versus a variable number of line outages for both MDL and sample variance tests.

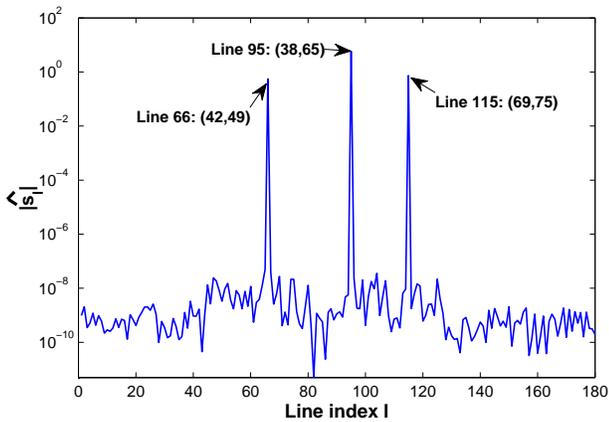

Fig. 3. Absolute value of $\hat{s}_\ell^\lambda$ versus the line index $\ell$, for the $\lambda$ value yielding three line outages.

## VI. CONCLUSIONS AND CURRENT RESEARCH

A novel line-change identification algorithm was developed in this paper for power system monitoring, through an overcomplete representation capturing the sparse innovation manifested by line changes on phasor angle measurements. This new formulation allows identification of multiple (even cascaded) transmission line outages, by regularizing an LS criterion with the $\ell_1$-norm of the sparse vector comprising the overcomplete representation parameters. As a result, the combinatorially complex line outage identification problem is converted to a computationally tractable convex QP. Thanks to its simple closed-form updates, the resultant BCD-based near real-time solver can efficiently compute the entire line-change identification path for any prescribed degree of sparsity. Practical statistical tests were also adapted to identify the actual number of line outages. Numerical tests demonstrated the merits of the proposed scheme in unveiling multiple line outages occurring even at the unobservable external system.

The proposed algorithm is currently evaluated on more complicated power testing systems, and various cases of line outages. Further enhancements to its accuracy are pursued by incorporating information on the injected power at the internal system, and also by extending it to cope with network faults other than line outages, e.g., changes in transmission line parameters. At the same time, the novel overcomplete representation for line outages is tested in the context of the *nonlinear* AC power flow model, which has wider applicability in practice.